\documentclass[12pt,a4paper,twoside]{article}
\usepackage{epsfig}

\usepackage{array,float,graphicx,amssymb,fancybox}
\usepackage{amsmath,bm,color}
\usepackage{fancyhdr}
\usepackage{lastpage}
\title{ The incremental bulk modulus, Young's modulus and Poisson's ratio in nonlinear isotropic elasticity:  physically reasonable response }
\author{N.H. Scott\thanks{Email: n.scott@uea.ac.uk}\\
School of Mathematics, University of East Anglia,\\
Norwich Research Park,  Norwich NR4 7TJ, U.K.}

\date{12 May 2020}

\begin{document}
\maketitle

\thispagestyle{fancy} \lhead{\emph{Mathematics and Mechanics of Solids}  (2007) {\bf 12}, 526--542.\\ 
doi: 10.1177/1081286506064719\\ 
Published 7 April 2006
}
\rhead{Page\ \thepage\ of\ \pageref{LastPage}}
\cfoot{}

\begin{center}
\emph{[Dedicated to Michael Hayes with esteem and gratitude]}
\end{center}

\begin{abstract}
An incremental (or tangent) bulk modulus for finite isotropic elasticity is defined which compares an increment in hydrostatic pressure with the corresponding increment in relative volume.  Its positivity provides a stringent criterion for physically reasonable response involving the second derivatives of the strain energy function.  Also, an average (or secant) bulk modulus is defined by comparing the current stress with the relative volume change.  The positivity of this bulk modulus provides a physically reasonable response criterion less stringent than the former.  The concept of incremental bulk modulus is extended to anisotropic elasticity.  For states of uniaxial tension an incremental Poisson's ratio and an incremental Young's modulus  are similarly defined  for nonlinear isotropic elasticity and have properties similar to those of the incremental bulk modulus.  The incremental Poisson's ratios for the isotropic constraints of incompressibility, Bell, Ericksen, and constant area are considered.  The incremental moduli are all evaluated for a specific example of the compressible neo-Hookean solid.   Bounds on the ground state Lam\'{e} elastic moduli, assumed positive, are given which are sufficient to guarantee the positivity of the incremental bulk and Young's moduli for all strains.  However, although the ground state Poisson's ratio is positive we find that the incremental Poisson's ratio becomes negative for large enough axial extensions.

\noindent{\bf Keywords} Nonlinear elasticity, incremental elastic moduli, compressible neo-Hookean material, constant-area constraint, superficial incompressibility\\
{\bf MSC (2010)} 74B20\\
\end{abstract}

\section{Basic nonlinear and linear isotropic elasticity}
\setcounter{equation}{0}

In this section we recount the basic theory of nonlinear and linear elasticity which we shall need before defining in Section 2 the incremental bulk and Young's moduli and the incremental Poisson's ratio and deriving some of their properties.  In Section 3 we consider these moduli in detail for a specific example of the compressible neo-Hookean material.

\pagestyle{fancy}
\fancyhead{}
\fancyhead[RO,LE]{Page\ \thepage\ of\ \pageref{LastPage}}
\fancyhead[LO]{Incremental  moduli in nonlinear elasticity}
\fancyhead[RE]{N. H. Scott}

In terms of the deformation gradient $\bf F$ the left and right Cauchy-Green tensors are $\bf B=FF^{\rm T}$ and $\bf C=F^{\rm T}F$, respectively.  In terms of the three invariants 
\begin{equation}
J_1={\rm tr\,}{\bf C},\;\;\; J_2={\rm tr\,}{\bf C}^{-1},\;\;\; J=\sqrt{\det{\bf C}}
\label{1.1}
\end{equation}
the Cauchy stress of an isotropic elastic material may be written
\begin{equation}
\mbox{\boldmath $\sigma$}=\beta_0{\bf I}+\beta_1{\bf B}+\beta_{-1}{\bf B}^{-1}
\label{1.2}
\end{equation}
in which $\bf I$ denotes the unit tensor and $\beta_\Gamma$ $(\Gamma=0,1,-1)$ are functions only of the three invariants $J_1,\,J_2,\,J$ defined at (\ref{1.1}).  The response functions $\beta_\Gamma$ are often assumed to satisfy the  empirical inequalities \cite[(51.27)]{1}
\begin{equation}
\beta_0\leq0,\;\;\; \beta_1>0,\;\;\;\beta_{-1}\leq0.
\label{1.3}
\end{equation}
It seems reasonable to assume that these inequalities are satisfied for small enough strains of a non-linearly elastic material.  In the theoretical discussion of Section 2 the inequalities (\ref{1.3}) are assumed at the beginning of Sections 2.1 and 2.2 in order to show that equal principal stresses imply equal corresponding principal stretches.  In fact, all that is required is the invertibility of the stress-strain law (\ref{1.2}).  For the specific example of a certain compressible neo-Hookean material considered in Section 3 it is shown that the empirical inequalities (\ref{1.3}) are indeed satisfied for deformations for which the invariant $J$ lies within a certain range of values, see (\ref{3.2a}) below.  Other possible restrictions upon the response functions are considered in \cite{1}.

If the isotropic elastic material is hyperelastic there exists a strain-energy function $W(J_1,J_2,J)$ such that the Cauchy stress (\ref{1.2}) reduces to
\begin{equation}
\mbox{\boldmath $\sigma$}=W_J{\bf I}+2J^{-1}W_1{\bf B} - 2J^{-1}W_2{\bf B}^{-1}
\label{2.1}
\end{equation}
in which $W_1:=\partial W/\partial J_1$, $W_2:=\partial W/\partial J_2$ and $W_J:=\partial W/\partial J$.  Comparing with (\ref{1.2}) we see that
\begin{equation}
 \beta_0=W_J,\;\;\; \beta_1=2J^{-1}W_1,\;\;\; \beta_{-1}=-2J^{-1}W_2
\label{2.2}
\end{equation}
and the empirical inequalities (\ref{1.3}) become
\begin{equation}
W_J\leq 0,\;\;\;  W_1 >0,\;\;\; W_2\geq 0.
\label{2.3}
\end{equation}

From (\ref{2.1}) the principal Cauchy stresses $\sigma_{ii}$ $(i=1,2,3)$ are given in terms of the principal stretches $\lambda_i$ $(i=1,2,3)$ by 
\begin{equation}
\sigma_{ii}=W_J+2J^{-1}W_1\lambda_i^2-2J^{-1}W_2\lambda_i^{-2}\;\;\; (i=1,2,3).
\label{2.4}
\end{equation}

In terms of the Lam\'{e} moduli $\lambda_0$, $\mu_0$ of linear isotropic elasticity, the bulk modulus, Poisson's ratio and Young's modulus are given by
\begin{equation}
K_0=\lambda_0+{\textstyle\frac23}\mu_0,\;\;\; \nu_0=\frac{\lambda_0}{2(\lambda_0+\mu_0)},\;\;\; E_0=\frac{\mu_0(3\lambda_0+2\mu_0)}{\lambda_0+\mu_0},
\label{2.5}
\end{equation}
see \cite[Table 3]{2}.  The positive definiteness of the strain energy of linear isotropic elasticity requires the inequalities
\begin{equation}
\mu_0 > 0, \;\;\; K_0>0
\label{3.1}
\end{equation}
to hold.  In turn these inequalities imply the further inequalities
\begin{equation}
-1<\nu_0<{\textstyle\frac12},\;\;\; E_0>0,\;\;\; \lambda_0> - {\textstyle\frac23}\mu_0.
\label{3.2}
\end{equation}
Commonly in linear isotropic elasticity it is found that inequalities more restrictive than (\ref{3.2}) hold:
\begin{equation}
0<\nu_0<{\textstyle\frac12},\;\;\; E_0>0,\;\;\; \lambda_0> 0.
\label{3.3}
\end{equation}

\section{ Incremental elastic moduli}
\setcounter{equation}{0}
\subsection{The incremental bulk modulus}

Consider an initially stress-free isotropic elastic material subject to a hydrostatic pressure $p=-\sigma_{11} = -\sigma_{22}= -\sigma_{33} = -\sigma_{pp}/3$ given by (\ref{1.2}) and (\ref{2.4}), employing the summation convention on repeated suffixes.  By subtraction of the diagonal components of (\ref{1.2}) and use of (\ref{1.3}$)_{2,3}$ it can be shown that the principal stretches must be equal, so that $\lambda_1= \lambda_2=\lambda_3=\lambda$, say,  and therefore the strain resulting from the pure hydrostatic pressure $\mbox{\boldmath $\sigma$}= -p{\bf I}$ is the pure dilatation $\bf F=\lambda I$.  After the application of the hydrostatic pressure $p$, an initial volume $V_0$ becomes $V=V_0J=V_0\lambda^3$.  A further increment $\delta p$ in pressure results in a further volume change $\delta V=3V_0\lambda^2\delta\lambda$.  The incremental bulk modulus is defined by
\[\mbox{Incremental bulk modulus} = -\, \frac{\mbox{change in pressure}}{\mbox{relative change in volume}}
\]
\begin{equation}
=\frac{-\delta p}{\delta V/V} = \frac{\delta\sigma_{11}}{\delta(V_0\lambda^3)/(V_0\lambda^3)} = 
\frac{J\delta\sigma_{11}}{\delta J} =
\frac{\lambda\delta\sigma_{11}}{3\delta\lambda}.
\label{4.1}
\end{equation}
Thus the incremental (or tangent) bulk modulus is defined in the limit $\delta\lambda\rightarrow 0$ by either of the equivalent forms
\begin{equation}
K(\lambda):= \frac{\lambda}{3}\left( \frac{\partial\sigma_{11}}{\partial\lambda}\right)_{\lambda_1=\lambda_2=
\lambda_3=\lambda},
\label{5.1}
\end{equation}
or
\begin{equation}
K(\lambda):= \frac{J}{3}\left( \frac{\partial\sigma_{pp}}{\partial J}\right)_{\lambda_1=\lambda_2=
\lambda_3=J^{1/3}}.
\label{5.1a}
\end{equation}

For a pure dilatation of stretch $\lambda$ we have $J_1=3\lambda^2$, $J_2=3\lambda^{-2}$, $J=\lambda^3$ and (\ref{2.4}) reduces to
\begin{equation}
\sigma_{11}=W_J+2\lambda^{-1}W_1-2\lambda^{-5}W_2.
\label{5.2}
\end{equation}
Use of (\ref{5.2}) in (\ref{5.1}) shows that the incremental bulk modulus of isotropic hyperelasticity may be expressed in the form
\begin{equation}
K(\lambda)= - {\textstyle\frac23}\lambda^{-1}W_1 +
{\textstyle\frac{10}{3}}\lambda^{-5}W_2 \hspace*{69mm}
\label{5.3}
\end{equation}
\[
\hspace*{20mm} \mbox{}+\lambda^3W_{JJ} +4\lambda W_{11}+4\lambda^{-7}W_{22}+4\lambda^2W_{J1}-4\lambda^{-2}W_{J2}-8\lambda^{-2}W_{12}
\]
in which $W_{12}$ denotes $\partial^2W/\partial \lambda_1\partial \lambda_2$, etc.  This is similar to the incremental bulk moduli of Rivlin and Beatty \cite{R&B} in their analysis of the stability of a compressible unit cube under dead loading.  Rivlin and Beatty \cite{R&B} also define incremental Young's moduli and Poisson's ratios similar to those defined below.  These incremental elastic moduli are also related to the generalized Lam\'{e} moduli of Beatty 
\cite[Appendix]{B1}.

On physical grounds we might expect $\delta p$ and $\delta V$ to have opposite signs, so that, for example, a further increase in pressure results in a further decrease in volume.  Then physically reasonable response would require
\begin{equation}
K(\lambda)>0\;\;\;  \mbox{for}\;\;\;  0<\lambda<\infty,
\label{5.4}
\end{equation}
 a criterion  which may be shown to be equivalent to the P-C inequality of Truesdell and Noll \cite[(51.3)]{1}.   However, for an empirically determined material it might not be possible to insist that $K(\lambda)$ be positive for all $\lambda$.  We might find that $K(\lambda)$ can be positive only for a smaller $\lambda$-interval, provided that this interval includes the stress-free ground state $\lambda = 1$.  Then the material model is physically realistic only for strains within this $\lambda$-interval.   We find in the limit $\lambda\rightarrow 1$ of (\ref{5.1}) that $K(1)=K_0$, so that the ground state incremental bulk modulus is equal to the usual bulk modulus of linear isotropic elasticity.

We may define an  average (or secant) bulk modulus in which the current stress is compared with the relative volume change:
\begin{equation}
\overline{K}(\lambda):= \frac{\sigma_{11}}{\lambda^3-1}.
\label{7.1}
\end{equation}
We find that $\overline{K}(1)=K_0=K(1)$, so that this alternative definition of bulk modulus agrees with (\ref{5.1}) in the ground state.  Physically reasonable response would require
\begin{equation}
\overline{K}(\lambda)>0\;\;\;  \mbox{for}\;\;\;  0<\lambda<\infty.
\label{7.2}
\end{equation}
However, this condition is less restrictive than (\ref{5.4}) as it requires only that $\sigma_{11}$ and $\lambda-1$ have the same sign, not that $\delta\sigma_{11}$ and $\delta\lambda$ should have the same sign.  

By integrating (\ref{5.1}) we find that, since the reference configuration $\lambda=1$ is stress free, (\ref{7.1}) may be written
\begin{equation}
\overline{K}(\lambda) =\frac{3}{\lambda^2+\lambda+1} \cdot
\frac{1}{\lambda-1}\int^\lambda_1 \frac{K(\lambda^\prime)}{\lambda^\prime}\,{\rm d}\lambda^\prime,
\label{7.3}
\end{equation}
clarifying the sense in which  $\overline{K}(\lambda)$ may be regarded as an average of the incremental bulk modulus.  It is clear from (\ref{7.3}) that criterion (\ref{5.4}) implies (\ref{7.2}) but that the converse does not hold: it is possible for $K(\lambda)$ to be negative with $\overline{K}(\lambda)$ remaining positive, see, for example, (\ref{17.2}) and (\ref{18.4}).

That (\ref{5.4}) is a more refined criterion for physically reasonable response than (\ref{7.2}) is not surprising when we note from (\ref{5.3}) that $K(\lambda)$ involves the second derivatives of the strain energy function $W$ whereas $\overline{K}(\lambda)$ involves only the first derivatives, see (\ref{5.2}) and (\ref{7.1}).

The notion of incremental bulk modulus may be extended to anisotropic nonlinear elasticity as follows.  Define a modified deformation gradient by ${\bf F}^{\displaystyle *} = J^{-\frac13}{\bf F}$ so that $\det{\bf F}^{\displaystyle *}\equiv 1$ for all deformations.  The modified right Cauchy-Green tensor is
\begin{equation}
{\bf C}^{\displaystyle *} = {{\bf F}^{\displaystyle *}}^{\rm T}  {\bf F}^{\displaystyle *} = J^{-\frac23}{ \bf C}\;\;\;\mbox{which  satisfies}\;\;\;
 \det {\bf C}^{\displaystyle *} \equiv 1.
\label{6.1}
\end{equation}
Then pure dilatation is ${\bf C}^{\displaystyle *} = {\bf I}$, $J\neq 1$ and pure distortion is  ${\bf C}^{\displaystyle *} \neq {\bf I}$, $J = 1$.
 The strain energy may be written
\begin{equation}
W^{\displaystyle *}({\bf C}^{\displaystyle *},\,J) \equiv W({\bf C}).
\label{6.2}
\end{equation}
From \cite[(84.11)]{1}, the Cauchy stress is
\begin{equation}
\sigma_{ij} =   W^{\displaystyle *}_J\delta_{ij}-{\textstyle\frac23}  J^{-1}C^{\displaystyle *}_{CD}\frac{\partial  W^{\displaystyle *}}
{\partial  C^{\displaystyle *}_{CD}}\delta_{ij}  + J^{-1}\left(
F^{\displaystyle *}_{iC}F^{\displaystyle *}_{jD}+ F^{\displaystyle *}_{iD} F^{\displaystyle *}_{jC}\right)\frac{\partial  W^{\displaystyle *}}
{\partial  C^{\displaystyle *}_{CD}}.
\label{6.4}
\end{equation}
Taking the trace gives simply $\sigma_{pp}=3W^{\displaystyle *}_J$ so that from the definition (\ref{5.1a}) most suited to anisotropic elasticity the anisotropic incremental bulk modulus~is
\begin{equation}
K ({\bf C}^{\displaystyle *},\, J) := \frac{J}{3}\frac{\partial\sigma_{pp}}{\partial J} = JW^{\displaystyle *}_{JJ}.
\label{6.5}
\end{equation}

Ogden \cite[(7.4.39)]{3} and Scott \cite[(4.7)]{4} omit the factor $J$ in (\ref{6.5})  because they compare the  volume change with the original volume $V_0$ rather than with the current volume $V=V_0J$.

\subsection{The incremental Poisson's ratio}

Our initially stress free isotropic material is subjected to a uniaxial tension $\sigma_{11}$ with all other stress components vanishing.  The corresponding axial principal stretch is denoted by $\lambda_1$ and the lateral principal stretches are $\lambda_2=\lambda_3$.  That these are equal in the uniaxial tension of an isotropic material follows from (\ref{1.3}), see Batra \cite{5}.  Then the invariants (\ref{1.1}) are given by
\begin{equation}
J_1=\lambda_1^2+2\lambda_2^2,\;\;\; J_2=\lambda_1^{-2}+2\lambda_2^{-2},\;\;\; J=\lambda_1\lambda_2^2
\label{8.1}
\end{equation}
and the vanishing of the lateral stresses $\sigma_{22}$ and $\sigma_{33}$ requires, from (\ref{2.1}),
\begin{equation}
\sigma_{22}=W_J+2\lambda_1^{-1}W_1-2\lambda_1^{-1}\lambda_2^{-4}W_2=0.
\label{8.2}
\end{equation}
For a given strain energy $W$ this equation gives an implicit relation between $\lambda_1$ and  $\lambda_2$, so that
\begin{equation}
\lambda_2=\lambda_2(\lambda_1)
\label{8.3}
\end{equation}
may be regarded as a known function.

After application of the uniaxial stress $\sigma_{11}$ a cylinder of original length $L$ and radius $R$, with generators parallel to the 1-axis, becomes a cylinder of length $\lambda_1L$ and radius $\lambda_2R$ with the same generators.  The application of a further uniaxial stress $\delta\sigma_{11}$ (maintaining $\sigma_{22}=0$) leads to changes $\delta\lambda_1$ in $\lambda_1$ and $\delta\lambda_2$ in $\lambda_2$.  We may define an incremental Poisson's ratio by
\[
\mbox{Incremental Poisson's ratio}= -\, \frac{\mbox{relative change in radius}}{\mbox{relative change in length}}
\]
\begin{equation}
=  -\, \frac{\delta(\lambda_2R)/(\lambda_2R)}{\delta(\lambda_1L)/(\lambda_1L)}
=-\, \frac{\delta\lambda_2/\lambda_2}{\delta\lambda_1/\lambda_1}
= - \,\frac{\lambda_1}{\lambda_2}\frac{{\delta\lambda_2}}{{\delta\lambda_1}}.
\label{9.1}
\end{equation}
Thus we define the incremental Poisson's ratio by taking the limit $\delta\lambda_1\rightarrow 0$:
\begin{equation}
\nu(\lambda_1):= - \,\frac{\lambda_1}{\lambda_2}\left(
\frac{\partial\lambda_2}{\partial\lambda_1}
\right)_{\lambda_2=\lambda_3,\; \sigma_{22}=\sigma_{33}=0}
= -\, \frac{\lambda_1}{\lambda_2}\frac{{\rm d}\lambda_2}{{\rm d}\lambda_1},
\label{9.2}
\end{equation}
in which the first expression is obtained by differentiating (\ref{8.2}) implicitly and the second, equivalently, by differentiating (\ref{8.3}) explicitly.

We commonly expect an increase $\delta\lambda_1$ in axial stretch to be accompanied by a decrease $\delta\lambda_2$ in lateral stretch and a decrease  $\delta\lambda_1$ to be accompanied by an increase $\delta\lambda_2$.  In either case, $\delta\lambda_1$ and $\delta\lambda_2$ have opposite sign and Poisson's ratio defined by (\ref{9.2})  is positive.  However, it can be seen from (\ref{3.2}$)_1$ in the linear case that
\begin{equation}
\nu(\lambda_1) > 0
\label{10.1}
\end{equation}
cannot be a universal requirement of physically reasonable response in the way that, for example, $K(\lambda)>0$ is for the bulk modulus, see (\ref{5.4}).  On taking the limit $\lambda_1\rightarrow 1$ we see that $\nu(1)=\nu_0$, so that the ground state incremental Poisson's ratio is equal to the usual Poisson's ratio of linear isotropic elasticity.

In some circumstances we find that (\ref{8.3}) takes the simple form
\begin{equation}
\lambda_2=\lambda_1^{-\nu_0}
\label{11.1}
\end{equation}
in which $\nu_0$ is a constant.  Then from (\ref{9.2}) the incremental Poisson's ratio takes the value
\begin{equation}
\nu(\lambda_1)=\nu_0,
\label{11.2}
\end{equation}
constant for all values of the axial stretch $\lambda_1$.  Conversely, if (\ref{11.2}) holds then (\ref{9.2}) may be integrated to recover (\ref{11.1}) if it is remembered that $\lambda_1=\lambda_2=1$ in the reference configuration.

Beatty and Stalnaker \cite[(2.5)]{6}  introduce a function which they term the Poisson function of finite elasticity defined by
\begin{equation}
\overline{\nu}(\lambda_1)= - \,\frac{\lambda_2-1}{\lambda_1-1}
\label{11.4}
\end{equation}
in which $\lambda_2(\lambda_1)$ is given implicitly by (\ref{8.2}).  This is a (secant)  form of Poisson's ratio in which the  overall lateral  extension is compared with the   overall axial  extension.  As with $\nu(\lambda_1)$, we find that $\overline{\nu}(1)=\nu_0$, agreeing with the ground state Poisson's ratio.  By integrating (\ref{9.2}) we find that Beatty and Stalnaker's Poisson function is given by
\begin{equation}
\overline{\nu}(\lambda_1)=\frac{1}{\lambda_1-1}
\int^{\lambda_1}_1\frac{\lambda_2(\lambda_1^\prime)\nu(\lambda_1^\prime)}
{\lambda_1^\prime}\,{\rm d}\lambda_1^\prime
\label{11a}
\end{equation}
and so   may be regarded as an average of  the incremental Poisson's ratio $\nu(\lambda_1)$.  If $\nu(\lambda_1)$ does not change sign on an interval containing $\lambda_1=1$ then (\ref{11a}) shows that $\overline{\nu}(\lambda_1)$ bears the same sign on that interval.  But it is possible for $\nu(\lambda_1)$ to change sign on an interval without $\overline{\nu}(\lambda_1)$ doing so.  It is easy to see that $\nu(\lambda_1)$ involves second derivatives of $W$ whilst $\overline{\nu}(\lambda_1)$ involves only first derivatives.  Clearly, $\nu(\lambda_1)$ and $\overline{\nu}(\lambda_1)$ bear a similar relationship to each other as do $K(\lambda)$ and $\overline{K}(\lambda)$.

\subsubsection*{Constrained materials}

The four constraints considered in this subsection are {\em isotropic}, i.e. the form of the constraint is  invariant under any interchange of principal stretches $\lambda_i$.  Such constraints are the most natural to consider in an isotropic material.

\paragraph{Incompressibility.} For an  incompressible material all motions are isochoric so $J=\lambda_1\lambda_2^2=1$.  Then $\lambda_2=\lambda_1^{-1/2}$, which is of the form (\ref{11.1}), so the incremental Poisson's ratio for an incompressible isotropic material is
\begin{equation}
\nu(\lambda_1)=\frac12
\label{11.3}
\end{equation}
for all values of $\lambda_1$.  This is the well known value of Poisson's ratio for  incompressible linear  isotropic elasticity.  For incompressibility Beatty and Stalnaker's function becomes
\begin{equation}
\overline{\nu}(\lambda_1)=\frac{1}{\lambda_1+\lambda_1^{1/2}},
\label{12.1}
\end{equation}
dependent on $\lambda_1$,  though in the ground state $\lambda_1=1$  this also  reduces to the value $\overline{\nu}=\nu=1/2$ of incompressible linear isotropic elasticity.   

\paragraph{Bell's constraint.} In many experiments on metal plasticity Bell \cite{7} found the constraint
\begin{equation}
 \lambda_1+\lambda_2+\lambda_3=3
\label{12.3}
\end{equation}
to hold.  With $\lambda_2=\lambda_3$ the two definitions of Poisson's ratio give
\begin{equation}
\nu(\lambda_1)=\frac{\lambda_1}{3-\lambda_1},\;\;\; 
\overline{\nu}(\lambda_1)=\frac12,
\label{12.4}
\end{equation}
so that here Beatty and Stalnaker's definition is independent of $\lambda_1$, see \cite[(2.10)]{6}.   The limit $\lambda_1\rightarrow 3$ might be expected to be singular for a Bell material as it involves the limit $\lambda_2\rightarrow 0$
 and for (\ref{12.4}$)_1$ this is so, though not for  (\ref{12.4}$)_2$.   
 
\paragraph{Ericksen's constraint.} Ericksen \cite{8,9}, and later Scott \cite{10}, employed the constraint
\begin{equation}
 \lambda_1^2+\lambda_2^2+\lambda_3^2=3
\label{12.5}
\end{equation}
in a constitutive theory of elastic crystals.  The corresponding Poisson's ratios are
\begin{equation}
\nu(\lambda_1)=\frac{\lambda_1^2}{3-\lambda_1^2},\;\;\; 
\overline{\nu}(\lambda_1)=\frac{1+\lambda_1}{2+(6-2\lambda_1^2)^{1/2}}.
\label{13.1}
\end{equation}
The expected singularity of the  limit  $\lambda_1^2\rightarrow 3$ for Ericksen's constraint (\ref{12.5}) is borne out by (\ref{13.1}$)_1$ and also by (\ref{13.1}$)_2$ for which\footnote{Eqn (\ref{13.1}$)_2$ corrects an error in the original paper.} $\overline{\nu}(\lambda_1)$ becomes complex for $\lambda_1>\sqrt{3}$ and $\overline{\nu}(\sqrt{3})=\tfrac12 (1+\sqrt{3})$.

\paragraph{Constant-area constraint.} The   constant-area constraint, see  \cite{BoSa96},
\begin{equation}\label{area1}
\lambda_1\lambda_2 + \lambda_2\lambda_3 + \lambda_3\lambda_1 =3
\end{equation}
is so called because a material cube in the reference configuration with edges parallel to the principal strain axes retains the same total surface area after deformation.  The two corresponding Poisson's ratios are given by
\begin{equation}\label{area2}
\nu(\lambda_1) = \frac{\lambda_1}{\sqrt{\lambda_1^2 + 3}},\;\;\;
\overline{\nu}(\lambda_1) = \frac{2}{\sqrt{\lambda_1^2 + 3} + \lambda_1 + 1}.
\end{equation}

In the ground state all the Poisson's ratios (\ref{12.4}),  (\ref{13.1}) and (\ref{area2})  reduce to the value $1/2$ indicative of the fact that the constraints (\ref{12.3}),  (\ref{12.5}) and (\ref{area1}) are all equivalent to incompressibility for small  strains, as indeed are all isotropic constraints in this limit, see Destrade and Scott \cite[Section 3.2]{12}.

\subsubsection*{Interpretation of the ranges $\nu \geq \frac12$, $\nu \leq  -1$}

The results of this subsection do not depend on the material being elastic, only on the existence of a function of the form (\ref{9.2}) giving the lateral stretch as a function of the axial stretch in uniaxial tension.  These ranges are those normally excluded in linear isotropic elasticity, see (\ref{3.2})$_1$.

For any constant $n$, use of (\ref{9.2}) leads to
\begin{equation}\label{Int1}
\lambda_1\frac{{\rm d}}{{\rm d}\lambda_1}\left(\frac{\lambda_2}{\lambda_1^n}\right) = - \frac{\lambda_2}{\lambda_1^n}(\nu + n).
\end{equation}
Putting $n= -\frac12$ and remembering (\ref{8.1})$_3$  reduces   (\ref{Int1}) to
\begin{equation}\label{Int2}
\lambda_1\frac{{\rm d}}{{\rm d}\lambda_1}J^\frac12 =  - J^\frac12(\nu - \textstyle\frac12).
\end{equation}
If $\nu = \frac12$, independently of $\lambda_1$, then (\ref{Int2}) leads to $J=1$ so that uniaxial tension is isochoric as discussed at the beginning of the previous subsection.  If $\nu>\frac12$ for a range of $\lambda_1$-values including $\lambda_1=1$ then ${\rm d}J/{\rm d}\lambda_1 < 0$ leading to $J<1$.  Thus $\nu>\frac12$ implies that uniaxial tension leads to a volume decrease.  Similarly, $\nu<\frac12$ implies a volume increase.

Putting $n=  1$    reduces   (\ref{Int1}) to
\begin{equation}\label{Int3}
 \lambda_1\frac{{\rm d}}{{\rm d}\lambda_1}\left(\frac{\lambda_2}{\lambda_1}\right) = - \frac{\lambda_2}{\lambda_1}(\nu + 1).
\end{equation}
If $\nu =  - 1$, independently of $\lambda_1$, then (\ref{Int3})  implies that  $\lambda_2=\lambda_1$ so that uniaxial tension leads to a spherical deformation.  If  $\nu <  - 1$ then (\ref{Int3}) implies $\lambda_2>\lambda_1$ so that uniaxial tension leads to a lateral stretch greater than the axial stretch.  Thus a material sphere would become an oblate spheroid.  Similarly, if $\nu >  - 1$ then $\lambda_2<\lambda_1$ and a material sphere would become a prolate spheroid.

\subsection{The incremental Young's modulus}

An increment $\delta\sigma_{11}$ in the uniaxial tension $\sigma_{11}$ (applied whilst maintaining $\sigma_{22}=0$) of the isotropic elastic cylinder of the previous subsection is accompanied by an increment $\delta\lambda_1$ in the axial principal stretch $\lambda_1$.  We may define an incremental (or tangent) Young's modulus by
\[
\mbox{Incremental  Young's modulus}=  \frac{\mbox{ change in  uniaxial tension}}{\mbox{relative change in length}}
\]
\begin{equation}
=  \frac{\delta \sigma_{11}}{\delta(\lambda_1L)/(\lambda_1L)}
= \lambda_1 \frac{\delta \sigma_{11}}{\delta\lambda_1 }.
 \label{14.1}
\end{equation}
Thus we define the incremental  Young's modulus by taking the limit $\delta\lambda_1\rightarrow 0$:
\begin{equation}
E(\lambda_1):=  \lambda_1 \left(
\frac{\partial\sigma_{11}}{\partial\lambda_1}
\right)_{\lambda_2=\lambda_3,\; \sigma_{22}=\sigma_{33}=0}.
\label{14.2}
\end{equation}
From (\ref{2.1}), the uniaxial stress is given by
\begin{equation}
\sigma_{11}=W_J+2\lambda_1\lambda_2^{-2}W_1-2\lambda_1^{-4}\lambda_2^{-2}W_2
\label{14.3}
\end{equation}
and (\ref{14.2}) is obtained by differentiating (\ref{14.3}) implicitly bearing in mind (\ref{8.2}) and (\ref{8.3}).

On physical grounds we expect $\delta\sigma_{11}$ and $\delta\lambda_1$ to have the same sign, so that physically reasonable response would require
\begin{equation}
E(\lambda_1)>0\;\;\; \mbox{for}\;\;\; 0<\lambda_1<\infty,
\label{15.1}
\end{equation}
which accordingly we propose as a criterion for physically reasonable response, on a par with (\ref{5.4}).  The criterion (\ref{15.1}) is different from the tension-extension inequalities of Truesdell and Noll \cite[(51.10)]{1} because with the former the lateral stresses vanish whereas with the latter it is the lateral strains that vanish.  Some of the incremental moduli defined by Rivlin and Beatty \cite{R&B} have vanishing lateral strains,  rather than stresses, and so are not directly comparable with those defined here.  As with the bulk modulus, however, for an empirically determined material it may not be possible to insist that $E(\lambda_1)$ be positive for all $\lambda_1>0$.  We might find that $E(\lambda_1)$ can be positive only for a smaller $\lambda_1$-interval, provided it includes $\lambda_1=1$.  We have $E(1)=E_0$, so that the ground state incremental Young's modulus is equal to the usual Young's modulus of linear isotropic elasticity.

We may define an average (or secant) Young's modulus in which the uniaxial stress is compared with the relative axial extension:
\begin{equation}
\overline{E}(\lambda_1):=\frac{\sigma_{11}}{\lambda_1-1}
\label{15.2}
\end{equation}
and note that $\overline{E}(1)=E_0$, agreeing with (\ref{14.2}) evaluated in the ground state.  On physical grounds we might expect
\begin{equation}
\overline{E}(\lambda_1)>0\;\;\; \mbox{for}\;\;\; 0<\lambda_1<\infty,
\label{15.3}
\end{equation}
a condition less restrictive than (\ref{15.1}) as it requires only that $\sigma_{11}$ and $\lambda_1-1$ have the same sign, not that $\delta\sigma_{11}$ and $\delta\lambda_1$ should have the same sign.  
That (\ref{15.2}) is an average may be seen by integrating (\ref{14.2}) to obtain
\begin{equation}
\overline{E}(\lambda_1)=\frac{1}{\lambda_1-1}
\int^{\lambda_1}_1\frac{E(\lambda^\prime)}{\lambda^\prime}\,{\rm d}
\lambda^\prime.
\label{15.4}
\end{equation}
Clearly, if $E(\lambda_1)$ is positive on an interval containing $\lambda_1=1$ then so is $\overline{E}(\lambda_1)$ but $E(\lambda_1)$ could become negative without  $\overline{E}(\lambda_1)$ doing so.  Thus (\ref{15.1}) implies (\ref{15.3}) but not conversely.  $E(\lambda_1)$ involves second derivatives of $W$ whilst  $\overline{E}(\lambda_1)$ involves only first derivatives.  This situation is similar to that for bulk modulus and Poisson's ratio.

\section{ The compressible neo-Hookean solid: an example}
\setcounter{equation}{0}

We shall consider a specific form of the compressible neo-Hookean strain energy that has been considered previously by Ogden \cite[pp. 222--226]{3}:
\begin{equation}
W={\textstyle \frac12}\mu_0( J_1
-3 - 2\log J)+{\textstyle \frac12}\lambda_0(J-1)^2,
\label{16.2}
\end{equation}
in which $\lambda_0$ and $\mu_0$  are constants
 which we shall see may be identified with the ground state Lam\'{e} elastic moduli.  They are assumed to satisfy
\begin{equation}
\mu_0>0,\;\;\; \lambda_0>0.
\label{18.1}
\end{equation}
From (\ref{2.2}) we find that
\[\beta_0 = -\mu_0 J^{-1} + \lambda_0(J-1),\;\;\;
\beta_1 = \mu_0 J^{-1},\;\;\; \beta_{-1} = 0,\]
so that the empirical inequalities (\ref{1.3})$_{2,3}$ are clearly satisfied.  With (\ref{18.1}) holding, the remaining empirical inequality (\ref{1.3})$_1$ is satisfied provided that the invariant $J$ lies in the range
\begin{equation}\label{3.2a}
0<J<J_{\rm max}:={\textstyle\frac12}+{\textstyle\frac12}[1 + 4\mu_0/\lambda_0]^{1/2},
\end{equation}
both extremes of which are unphysical since $J_{\rm max}$  is attainable only in the unphysical limit $\lambda_2\rightarrow 0$,  $\lambda_1\rightarrow \infty$.
This range includes the value $J=1$ corresponding to the undeformed material.
We shall see that this allowable $J$-range  is relevant also to our subsequent discussion of Poisson's ratio and Young's modulus.

 The principal Cauchy stresses are obtained from (\ref{2.4})
 and (\ref{16.2}):
\begin{equation}
\sigma_{ii}=\mu_0J^{-1}(\lambda_i^2-1) + \lambda_0(J-1),\;\;\; 
i=1,2,3. 
\label{17.1}
\end{equation}

\subsection{Bulk modulus}

If the material (\ref{16.2}) is subjected to a uniform dilatation of stretch $\lambda$ each principal stress is equal to
\begin{equation}
\sigma_{11}=\mu_0(\lambda^{-1} - \lambda^{-3})+
\lambda_0(\lambda^{3}-1)
\label{17.a}
\end{equation}
and from (\ref{5.1}) the incremental bulk modulus is
\begin{equation}
K(\lambda)=\mu_0(\lambda^{-3} -{\textstyle \frac13} \lambda^{-1})+\lambda_0\lambda^{3}.
\label{17.2}
\end{equation}
Now
\[
K(1)=\lambda_0+{\textstyle \frac23}\mu_0=K_0
\]
the ground state bulk modulus, see (\ref{2.5}$)_1$.

We see that
\[K(\lambda) \sim \mu_0\lambda^{-3}\;\;(\lambda\rightarrow 0),\;\;\;
K(\lambda) \sim \lambda_0\lambda^{3}\;\;(\lambda\rightarrow \infty)
\]
and so the bulk modulus is positive at the extremes of the range of stretches provided that (\ref{18.1}) holds. 

We now investigate the condition on $\lambda_0$ and $\mu_0$ sufficient to force $K(\lambda)>0$ for all $\lambda>0$.  The derivative  $K^\prime(\lambda)$ is negative as $\lambda\rightarrow 0$ and positive as $\lambda\rightarrow \infty$ and so changes sign at least once on $(0,\infty)$ and this occurs when
\begin{equation}
9\lambda_0\mu_0^{-1}\lambda^6+\lambda^2-9=0.
\label{18.2}
\end{equation}
The left-hand side is monotonic on $(0,\infty)$  and so there is a unique root, say $\lambda_{\rm m}$, satisfying
\[
0<\lambda_{\rm m}<3.
\]
Since $\lambda_{\rm m}$ satisfies (\ref{18.2}) we find from (\ref{17.2}) that
\[
\mu_0^{-1}K(\lambda_{\rm m}) = 2\lambda_{\rm m}^{-3} - 
{\textstyle\frac49}\lambda_{\rm m}^{-1}.
\]
Then $K(\lambda_{\rm m})\geq 0$  provided $\lambda_{\rm m}
\leq3/\surd 2$.  Substituting this value of $\lambda$ into (\ref{18.2}) gives the condition sought:
\begin{equation}
\lambda_0\geq  {\textstyle\frac{4}{729}}\mu_0\;\;\; 
\Leftrightarrow\;\;\; K(\lambda)\geq 0\;\;\; \mbox{for}\;\;\; 0<\lambda<\infty.
\label{18.3}
\end{equation}
If the first inequality holds as an equality then the second inequality holds for all $\lambda>0$ except that it becomes an equality for the value $\lambda=3/\surd2$.  If the first inequality is violated then so is the second for a range of  positive values of $\lambda$.

The behaviour of the average bulk modulus $\overline{K}(\lambda)$ is simpler, and simpler to determine, but gives less physical insight.  On substituting the stress (\ref{17.a}) into the definition (\ref{7.1}) we find that for the material (\ref{16.2}) 
\begin{equation}
\overline{K}(\lambda) = \lambda_0 + 
\frac{\lambda+1}{\lambda^2+\lambda+1} \mu_0
\label{18.4}
\end{equation}
which is positive for all positive $\lambda$ provided that   the inequalities (\ref{18.1}) hold.  Since these inequalities are less restrictive than those (\ref{18.3}$)_1$ necessary to force $K(\lambda)$ to be positive we confirm that (\ref{5.4}) is a more stringent criterion for physically reasonable response than is (\ref{7.2}).  Nevertheless, if $\lambda_0<0$ then $\overline{K}(\lambda)$ becomes negative for large enough $\lambda$, as does $K(\lambda)$, see (\ref{17.2}).

\subsection{Poisson's ratio}

The material (\ref{16.2}) is subjected to the uniaxial tension $\sigma_{11}$ with corresponding axial stretch $\lambda_1$ and lateral stretches $\lambda_2=\lambda_3$.  Since $J=\lambda_1\lambda_2^2$,  the vanishing of the lateral stresses $\sigma_{22}=\sigma_{33}$ requires, from (\ref{17.1}),
\begin{equation}
\mu_0(1-\lambda_2^{-2})+\lambda_0\lambda_1(J-1)=0.
\label{19.1}
\end{equation}
Differentiating with respect to $\lambda_1$ and using the definition (\ref{9.2}) gives the expression
\begin{equation}
\nu(J)=\frac{(2J- 1)\lambda_0}
{2(J^{-1}\mu_0+J\lambda_0)}
\label{19.2}
\end{equation}
for the incremental Poisson's ratio.  Here $\nu$ is more conveniently expressed as a function of $J$ rather than $\lambda_1$.

In the ground state $\lambda_1=1,\; J=1$ and so
\[
\nu(1)=\frac{\lambda_0}{2(\mu_0+\lambda_0)}=\nu_0,
\]
the ground state Poisson's ratio, see (\ref{2.5}$)_2$.

We may multiply (\ref{19.1}) by $\lambda_2^2$ and obtain an equation quadratic in $J$
\begin{equation}
\lambda_0J^2-\lambda_0J+\mu_0(\lambda_2^2-1)=0,
\label{19.3}
\end{equation}
so that, provided the inequalities (\ref{18.1}) hold, $J$ has the two branches
\[
J_+={\textstyle\frac12}+{\textstyle\frac12}[1-\frac{4\mu_0}{\lambda_0}
(\lambda_2^2-1)]^{1/2},\;\;\; 0<\lambda_2^2 < 1+\frac{\lambda_0}{4\mu_0}
\]
and
\[
J_-={\textstyle\frac12}-{\textstyle\frac12}[1-\frac{4\mu_0}{\lambda_0}
(\lambda_2^2-1)]^{1/2},\;\;\; 1 <  \lambda_2^2 < 1+\frac {\lambda_0}{4\mu_0}
\]
viewed as  functions of $\lambda_2^2$.

It is clear from (\ref{19.2}) that $\nu>0$ on the upper branch $J_+$, $\nu<0$ on the lower branch $J_-$ and $\nu=0$ at the point $\lambda_2^2=1+\lambda_0/4\mu_0$, $J=1/2$,  where the two branches meet.  The point $\lambda_2=1, J=1$, of no deformation lies on $J_+$.  The allowable values of $J$ satisfy
\[
0<J<J_{\rm max},
\]
with $J_{\rm max}$ defined at (\ref{3.2a}).  Also from (\ref{3.2a}), we see that this is the same $J$-range as that prescribed by the empirical inequalities. 

If the inequalities (\ref{18.1}) hold it is clear from (\ref{19.1}) or (\ref{19.3}) that the following implications hold:
\[
\lambda_2^2<1 \Rightarrow J>1 \Rightarrow 
\lambda_1>\lambda_2^{-2}>1,
\]
\begin{equation}
\lambda_2^2>1 \Rightarrow J<1 \Rightarrow 
\lambda_1<\lambda_2^{-2}<1.
\label{21.1}
\end{equation} 
Thus in all circumstances $\lambda_2-1$ and $\lambda_1-1$ have opposite sign 
and we see that Beatty and Stalnaker's Poisson function (\ref{11.4}) satisfies $\overline{\nu} > 0$.    So for the material (\ref{16.2}) we always have  $\overline{\nu} > 0$, whereas for $J<1/2$ we have $\nu<0$.

\subsubsection*{Bounds on the value of $\nu$}

Poisson's ratio $\nu_0$ of linear isotropic elasticity is known to be bounded by (\ref{3.2}$)_1$ and we ask if there are corresponding bounds on the incremental Poisson's ratio $\nu$, given by (\ref{19.2}), of the compressible neo-Hookean material (\ref{16.2}).

We examine the behaviour of the incremental Poisson's ratio $\nu(J)$, given by (\ref{19.2}), on the interval $(0, J_{\rm max})$.  Clearly, $\nu(J)$ vanishes only at $J=0$ and $J=1/2$.  The derivative $\nu^\prime(J)$ vanishes on $J>0$ only for
\[J=J_{\rm min}:=\{(4\mu_0^2+\mu_0\lambda_0)^{1/2} - 2\mu_0\}/\lambda_0
\]
and for this value of $J$ the incremental Poisson's ratio attains its minimum value
\begin{equation}
\nu(J_{\rm min})=\nu_{\rm min}:=\frac{-\lambda_0}
{4\{(4\mu_0^2+\mu_0\lambda_0)^{1/2} + 2\mu_0\}},
\label{22.1}
\end{equation}
which is clearly negative.

Since $\nu^\prime(J)>0$ for $J>J_{\rm min}$ the largest value of $\nu$ occurs at $J=J_{\rm max}$ and is given by
\[
\nu(J_{\rm max})=\nu_{\rm max}:=\frac12,
\]
independently of $\lambda_0$ and $\mu_0$.

Then $\nu(J)$ is bounded by
\[
\nu_{\rm min}\leq \nu(J)\leq \frac12
\]
in place of (\ref{3.3}$)_1$ for linear isotropic elasticity with $\lambda_0>0$.  We have seen that $\nu_{\rm min}<0$ for $\lambda_0>0$.   If $ \lambda_0 = 32\mu_0 $ then $\nu_{\rm min}=-1$ and
\[
\lambda_0>32\mu_0 \;\Rightarrow\;  \nu_{\rm min}< -1.
\]
It is interesting to observe that in the limit $\lambda_0\rightarrow \infty$ of incompressibility we have $\nu_{\rm min} \rightarrow - \infty$,
though $\nu_0\rightarrow 1/2$.

From (\ref{18.3}) and (\ref{22.1}) we see that the inequalities
\begin{equation}
{\textstyle \frac{4}{729}}\mu_0 < \lambda_0 < 32\mu_0
\label{23.1}
\end{equation}
force $K(\lambda)>0$ and $-1<\nu(J)<\frac12$ for all $\lambda$ and relevant $J$.

\subsubsection*{The average Poisson's ratio $\overline{\nu}$}

From (\ref{19.3}) we may express $\lambda_2^2$ as a function of $J$ only:
\begin{equation}\label{prav1}
\lambda_2^2 = 1 + \frac{\lambda_0}{\mu_0}J(1 - J),
\end{equation}
so that $\lambda_1 = J/\lambda_2^2$ may also be expressed as a function of $J$ only.  Then the definition (\ref{11.4}) may be used to obtain $\overline{\nu}$ as a function of $J$:
\begin{equation}\label{prav2}
\overline{\nu}(J) = \frac{\left\{1 + \displaystyle\frac{\lambda_0}{\mu_0}J(1 - J)
\right\}\displaystyle\frac{\lambda_0}{\mu_0}J}
{\left\{\left[1 +  \displaystyle\frac{\lambda_0}{\mu_0}J(1 - J)\right]^\frac12 + 1 \right\}\left(1 + \displaystyle\frac{\lambda_0}{\mu_0}J\right)}.
\end{equation}
In the ground state $J=1$ we find that (\ref{prav2}) reduces to $\overline{\nu}(1) = \nu_0$ as expected, see (\ref{2.5}$)_2$.  $\overline{\nu}(J)$ is positive for all allowable values of $J$, vanishing only at the extremes of the $J$-range: $J=0$, $J=J_{\rm max}$.

\subsection{Young's modulus}

For any isotropic elastic material under uniaxial tension we see from (\ref{8.1}$)_3$ and the definition (\ref{9.2}) of the incremental Poisson's ratio that
\begin{equation}
\lambda_1\frac{{\rm d}J}{{\rm d}\lambda_1} = J(1-2\nu),
\label{23.2}
\end{equation}
equivalent to (\ref{Int2}).  
For the material (\ref{16.2}) we see from (\ref{17.1})  that the uniaxial stress is given by
\begin{equation}
\sigma_{11}=\mu_0J^{-1}(\lambda_1^2-1)+\lambda_0(J-1)
\label{23.3}
\end{equation}
and so from (\ref{23.2}) and the definition (\ref{14.2}) applied to (\ref{23.3}) we find that the incremental Young's modulus is given by
\begin{equation}
\mu_0^{-1}E(J)=\frac{J(1+2\nu)}{\left\{1 + \displaystyle\frac{\lambda_0}{\mu_0}J(1 - J)
\right\}^2} + 
\left(J^{-1}+ \frac{\lambda_0}{\mu_0}J\right)(1-2\nu),
\label{23.4}
\end{equation}
in which $\nu$ is given as a function of $J$ by (\ref{19.2}).   Again, $E$ is more conveniently expressed as a function of $J$ rather than $\lambda_1$.
In the unstressed reference configuration $J=1$, $\nu=\nu_0$ and we find that (\ref{23.4}) reduces to
\[
E(1) = \frac{\mu_0(3\lambda_0+2\mu_0)}{\lambda_0+\mu_0} = E_0,
\]
the ground state Young's modulus, see (\ref{2.5}$)_{2, 3}$.  We also see from (\ref{23.4})  that $E\rightarrow \infty$ as $J\rightarrow0$ or $J\rightarrow J_{\rm max}$.

The fact that the expressions for $K_0,$ $\nu_0$ and $E_0$ derived in this section all agree with the definitions (\ref{2.5}) confirms $\lambda_0$ and $\mu_0$ as the ground state Lam\'{e} moduli of the compressible neo-Hookean material (\ref{16.2}).

We have seen that $\nu<\frac12$ for this material and so the second term of (\ref{23.4}) is necessarily positive whereas if $\nu<-\frac12$ the first term is clearly negative.  However, from (\ref{22.1}) we see that 
$\lambda_0<12\mu_0 \;\Rightarrow\;  \nu_{\rm min}>-\frac12$ and it follows that
\[
\lambda_0<12\mu_0\;\; \;\Rightarrow\; \;\;
E(\lambda_1)>0\;\;\; \mbox{for}\;\;\; 0<\lambda_1<\infty.
\]
It is not known whether $E(\lambda_1)$ becomes negative for any larger values of $\lambda_0$, i.e. it is not known whether (\ref{15.1}) holds if (\ref{18.1}) does.  In order to say something about the Young's modulus we may replace the inequalities (\ref{23.1}) by the inequalities
\begin{equation}
{\textstyle \frac{4}{729}}\mu_0 < \lambda_0 < 12\mu_0
\label{23.6}
\end{equation}
which force $K >0$, $-\frac12 <\nu<\frac12$ and $E>0$.  

From (\ref{prav1}) and the definition (\ref{15.2}) we may derive an expression for the average Young's modulus in terms of $J$:
\begin{equation}\label{23.7}
\mu_0^{-1}\overline{E}(J) =\frac{1}{1 + \displaystyle\frac{\lambda_0}{\mu_0}J(1 - J)}
+ \frac{1}{J} + \displaystyle\frac{\lambda_0}{\mu_0} \cdot  \frac{1 + \displaystyle\frac{\lambda_0}{\mu_0}J(1 - J)}{1 + \displaystyle\frac{\lambda_0}{\mu_0}J}.
\end{equation}
In the ground state $J=1$ we find that (\ref{23.7}) reduces to $\overline{E}(1) = E_0$ as expected, see (\ref{2.5}$)_3$.  $\overline{E}(J)$ is positive for all allowable values of $J$ and, like $E$, we find that $\overline{E}\rightarrow\infty$ at the extremes of the $J$-range: $J=0$, $J=J_{\rm max}$.

The discussion of material (\ref{16.2}) has been entirely theoretical but we have seen that for quite large ranges of strain this material model behaves in accord with various notions of the physically reasonable response of a non-linearly elastic material.  The author has not, however, found any experimental results in the literature for materials that may be of the type  (\ref{16.2}).

\vspace{1cm}
\noindent{\it Acknowledgments.  I thank Professor Giuseppe Saccomandi for  inviting me to speak at Michael Hayes' retirement conference at Cortona, Italy in June, 2003 and the Istituto Nazionale di Alta Matematica, Roma, for generous financial support.}

\end{document}